# Magneto-Caloric effect and Multiple magnetic phases in Al doped $Ni_2MnSn_{0.75}Al_{0.25}$ Heusler Alloys

Satya Vijay Kumar[1$], Simran[2$], Madhusmita Jena[1], Mehroosh Fatema[1], Atul Gangwar[1], Srishti Dixit[1], Umashankar Rajput[1], Nisha Shahi[2], Chetna Gautam[3], Sanjay Singh[2], Anup K. Ghosh[3] and Sandip Chatterjee[1*]

## Affiliations:

[1] Department of Physics, Indian Institute of Technology (Banaras Hindu University), Varanasi 221005, Uttar Pradesh, India

[2] School of Materials Science and Technology, Indian Institute of Technology (Banaras Hindu University), Varanasi 221005, Uttar Pradesh, India

[3] Department of Physics, Banaras Hindu University, Varanasi 221005, Uttar Pradesh, India

*****Corresponding author: schatterji.app@iitbhu.ac.in

## Abstract

Among Heusler compounds, Ni-based alloys have been extensively investigated because they exhibit desirable properties such as high Curie temperatures, which are advantageous for advanced magnetic and spintronic devices.The effect of Al substitution on the magnetic ground state of $Ni_2MnSn$ was investigated using the $Ni_2MnSn_{0.75}Al_{0.25}$(NMSA) Heusler alloy. Temperature-dependent magnetisation measurements identify a second-order paramagnetic-to-ferromagnetic transition at $T_C \approx 734K$ , followed by a first-order martensitic transformation near 263 K, demonstrating strong magnetostructural coupling. Curie–Weiss analysis yields a positive Weiss temperature ($\theta_{CW}$=746.4 K) and an effective magnetic moment of 6.82 μB , confirming the predominance of ferromagnetic exchange interactions. The bifurcation between the ZFC and FCW magnetization curves, together with non-saturating hysteretic M–H loops, indicates the coexistence of competing ferromagnetic and antiferromagnetic interactions. Further magnetic investigations establish the formation of an interacting reentrant cluster-glass state accompanied by an exchange-bias effect. The observed magnetic behavior is attributed to the modification of Mn–Mn exchange interactions induced by Al substitution and the associated atomic disorder, resulting in a complex magnetic ground state.

## Introduction:

The development of multifunctional materials exhibiting coupled structural, magnetic, electronic, and thermal properties has attracted considerable attention because of their potential in energy conversion, sensing, and spintronic applications. Among the different classes of intermetallic compounds, Heusler alloys (HAs) have emerged as one of the most extensively investigated material systems owing to their rich physical properties and compositional versatility. Heusler alloys generally crystallize in the ordered $L2_1$ structure, and they may also form the partially ordered B2 structure, which significantly influences the electronic structure and magnetic exchange interactions [1-2]. Consequently, chemical substitution and atomic ordering provide effective routes for tailoring their multifunctional properties.[2]

Among the Heusler family, Ni–Mn-based alloys have attracted significant attention because of their rich magnetostructuralbehavior and multifunctional characteristics. These alloys exhibit ferromagnetism, martensitic transformation, magnetic shape-memory effect, magnetocaloric effect, and tunable electronic structure, making them attractive for both fundamental studies and technological applications [3-5]. Their magnetic properties originate primarily from exchange interactions between Mn atoms, while Ni mediates these interactions through hybridization. Since the magnetic exchange interaction is highly sensitive to Mn–Mn interatomic distance and atomic site occupancy, even small compositional modifications can significantly alter the magnetic ground state. This results in competing ferromagnetic and antiferromagnetic interactions simultaneously coexisting in these alloys, leading to complex magnetic behavior. Representative examples include$Pt_2MnGa$, $Ni_2MnAl$ ,[6], [7]and $Mn_2VSi$,[6] where antiferromagnetic ordering or competing magnetic interactions have been reported.[6-8].

This is of considerable interest for spintronics, largely because of the half-metallic character first described by de Groot and et.al. An ideal half-metal shows essentially 100% spin polarization, arising from a gap at the Fermi level in one of the two spin channels.[9] Because conduction in such half-metallic ferromagnets (HMFs) is carried almost entirely by spins of a single orientation, these materials are of interest for spintronic hardware tunnel magnetoresistance (TMR) sensors, spin valves, magnetic-field sensors, spin transistors, spin torque nano-oscillators (STNOs), and spin-polarised current injection into semiconductors. Full Heusler compounds in particular have attracted attention within this broader half-metallic family, due to their predicted half-metallicity, good structural match to common substrates, high spin polarization, sizable magnetic moments, shape-memory response, low magnetic damping, tunable anisotropy, and high Curie temperatures [9-12], all of which make them favorable candidates for producing spin-polarized current. Ni-Mn-based HAs specifically have been reported to show the exchange-bias effect, reentrant spin-glass behavior, memory and aging phenomena, and a giant MCE tied to their structural transitions [10,13]. Unlike ordinary magnetocaloric materials, which warm up under field because entropy drops, several Ni-Mn-In alloys instead cool down under an applied field an inverse effect that arises from how magnetic and structural entropy interact during the field-driven reverse martensitic transformation.

This cooling accompanies the martensite-to-austenite transition because the entropy gained structurally exceeds what is lost magnetically. A large share of the effort to strengthen the MCE and the related Magnetic Shape Memory Effect (MSME) in these alloys has gone toward enhancing magnetization in the austenite phase, most often by substituting Al or Co for Sn or Ni [9-14]. Liu et al., for example, reported that Co-doped $Ni_{45.2}Mn_{36.7}In_{13}Co_{5.1}$shows an adiabatic temperature change ($\Delta T_{ad}$) of 6.2 K at 317 K under a comparatively small 1.9 T field, and more broadly the Ni-Mn-In-(Co) family shows inverse-MCE $\Delta T_{ad}$values from 3.6 K to 6.2 K at 2 T[10]. Kihara and coworkers pushed this further, reaching $\Delta T_{ad}$= 12.8 K in $Ni_{45}Co_5Mn_{36.7}In_{13.3}$under a 15 T pulsed field [14],while Kai Numa et al. used the Maxwell relation to obtain a magnetic entropy change of 28.4 J/(kg K) in single-crystal $Ni_{45}Co_5Mn_{36.7}In_{13.3}$ at 292 K and 7 T [9]. Despite these remarkable characteristics, real-world use of Ni-Mn-In alloys is held back by thermal hysteresis, a limited working-temperature window, and irreversibility that stems from the first-order nature of the transition. Thermal cycling produces roughly 5-12 K of hysteresis in ternary Ni-Mn-In alloys, which widens to 10-30 K once Co is added [15]; as one example, moving from $Mn_{50}Ni_{39}In_9Co_2$to $Mn_{50}Ni_{37}In_9Co_4$extends the operating window from 24 K to 32 K, though at the expense of greater hysteresis.We also report the X-ray photoemission spectroscopy (XPS) results to investigate the electronic structure and chemical states of the constituent elements in the alloy.

## Result and Discussion

### 3.1 Rietveld Refined X-Ray Diffraction (XRD)

Figure 1 represents the Rietveld refined XRD pattern of $Ni_2MnSn_{0.75}Al_{0.25}$Heusler alloy, recorded at room temperature. The diffraction pattern observed is in good agreement with the calculated pattern, indicating good crystallinity of the Heusler phase [16-18]. The pattern obtained was refined considering it of cubic crystal structure and the reflections (111), (200), (220), (311), (222), (400), (422) and (511) were indexed. The appearance of the superlattice reflections (111) and (200) verifies the ordered Heusler structure of the alloy. Specifically, appearance of (111) reflection means that $L_{21}$ type ordering is present, and the intensity of this superlattice peak is relatively small, reflecting a small amount of atomic disorder caused by partial substitution of Al at the Sn site. The strongest diffraction peak, corresponding to the (220) plane, further confirms the formation of the cubic Heusler phase. The alloy crystallises mainly in cubic structure of the space group Fm-3m, which is typical for the full Heusler alloys[3]. Possible minor contribution from Pm-3m phase may also exist, resulting from local atomic disorder or antisite occupation caused by Al substitution. In the diffraction pattern, however, no significant impurity peak is observed, suggesting good phase purity of the sample prepared, as is also confirmed by the difference curve of the Rietveld refinement, which yields small difference peaks between the observed and calculated diffraction patterns. The sharp and narrow diffraction peaks denote good crystalline quality and long-range structural ordering of the alloy[19]. The difference in radius and electronic configuration of Sn and Al atoms slightly affects the lattice environment due to the substitution of Al into the Sn sublattice. The magnetic exchange interactions and/or the electronic band structure of the alloy can be modified by such structural modification, which in turn can change the physical properties of the alloy.

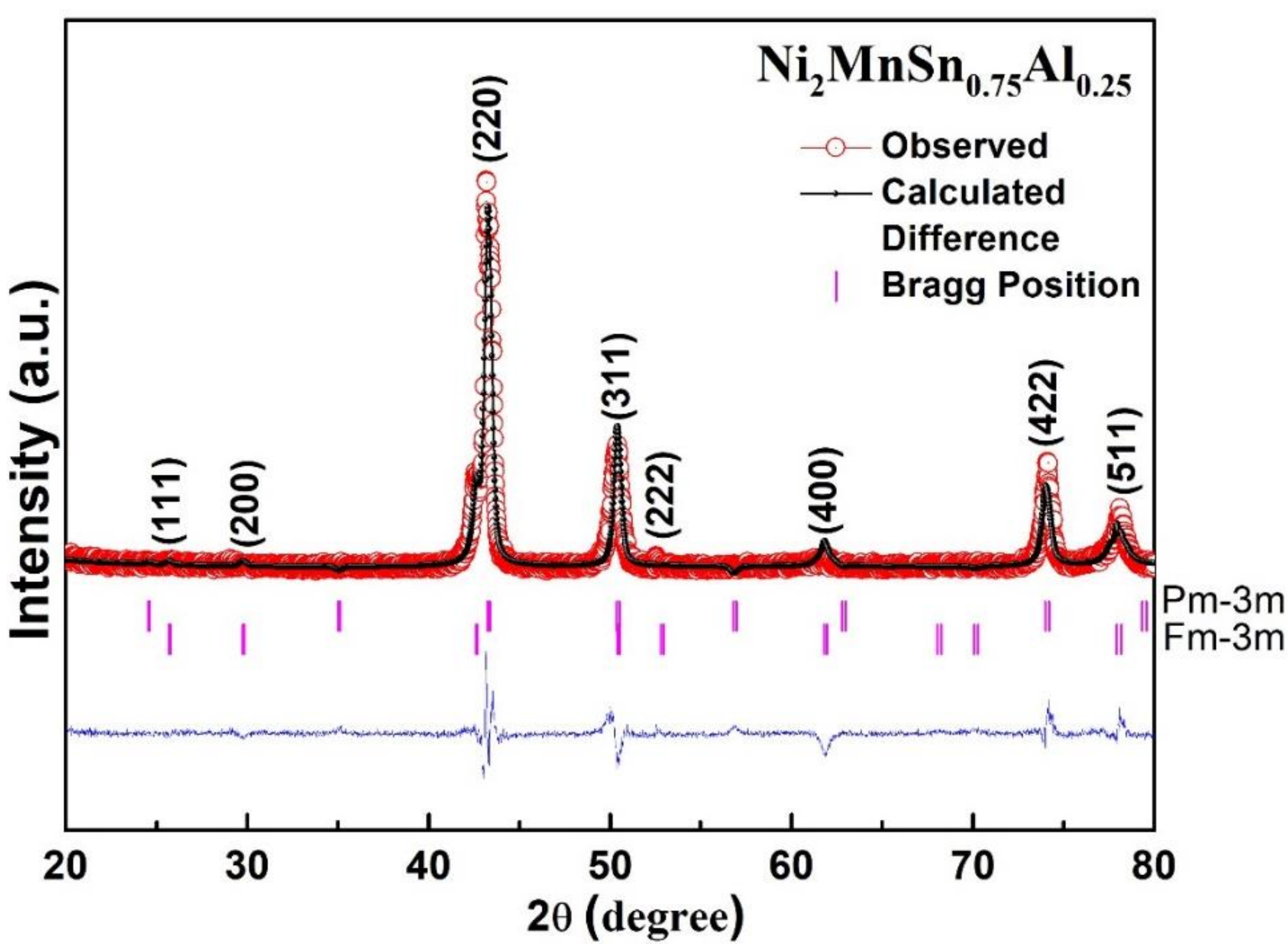

**Figure 1**: Rietveld Refined XRD pattern of $Ni_2MnSn_{0.75}Al_{0.25}$Heusler alloy.

**Microstructural and Compositional Analysis:**

In Figure 2(a), the HR-SEM image of the NMSAshows a smooth, homogeneous surface morphology. The elemental composition was further examined using energy-dispersive X-ray spectroscopy (EDX), which confirmed the presence of all constituent elements: Ni, Mn, Al, and Sn. All characteristic peaks corresponding to these elements are clearly observed, and no other peaks are detected, indicating the purity of the sample as shown in Figure 2(b). The estimated elemental concentrations closely match the nominal stoichiometric composition. Furthermore, the elemental mapping images as shown in Figs. 2(c)-2(f) reveal a uniform spatial distribution of all constituent elements.

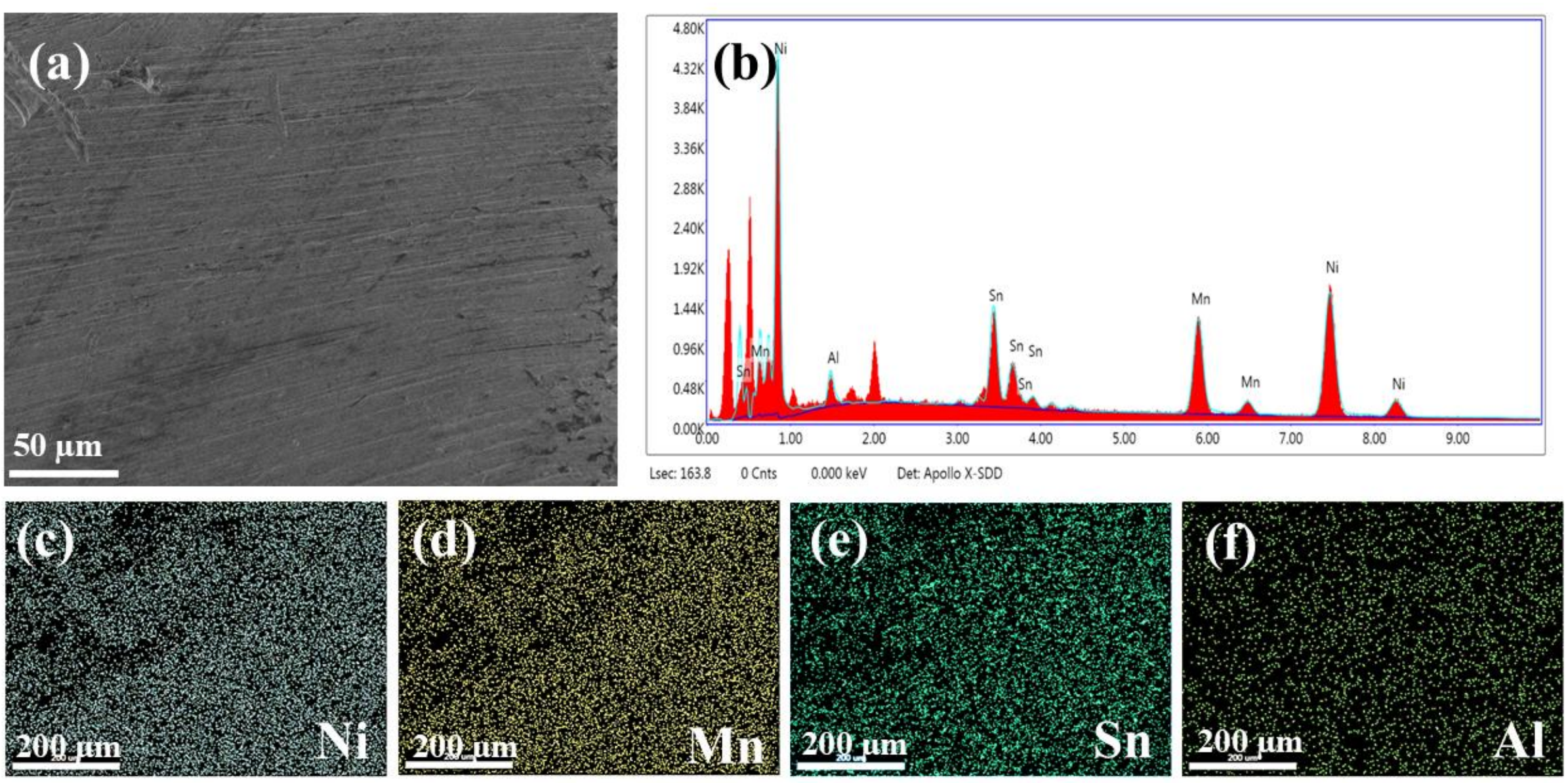

**Figure2**(a): FE-SEM image illustrating the surface morphology of the NMSA sample. (b) An energy-dispersive X-ray spectroscopy (EDX) spectrum is used to determine the elemental composition of the sample. (c) Elemental-mapping images showing the spatial distribution of the alloy's constituent elements in the NMSA alloy

## X-ray Photoelectron Spectroscopy (XPS) :

The high resolution XPS spectra of Ni 2p, Mn 2p, Sn 3d and Al 2p core levels obtained on the synthesized Heusler alloy $Ni_2MnSn_{0.75}Al_{0.25}$ are shown in Figure 3. The spectra can be deconvoluted, which provides information on the presence of all constituent elements and about their chemical state and the electronic environment in the alloy lattice[20]. Two characteristic spin-orbit split peaks corresponding to Ni $2p_{3/2}$ and Ni $2p_{1/2}$ are observed at 852-853 eV and 855-856 eV respectively in the Ni 2p spectrum[21]. The values of these binding energies are also similar to the values of the binding energies of Ni reported in Heusler compounds, suggesting Ni exists mainly in a metallic or weakly oxidized state ($Ni^{2+}$). The presence of a weak satellite feature at a high binding energy indicates some surface oxidation due to air exposure. The two peaks observed in the Mn 2p spectrum are around 640-641 eV and 652-653 eV, corresponding to the Mn $2p_{3/2}$ and the Mn $2p_{1/2}$ level, respectively*. The spin-orbit splitting observed shows good agreement with the ones reported in the Mn-based Heusler alloys. The Mn binding energies suggest a coexistence of $Mn^{2+}/Mn^{3+}$ states, and the broadened and asymmetric binding energy peak profile confirms that there is strong hybridization between Mn 3d states and neighbouring atoms, which is associated with magnetic exchange interactions within the alloy. The characteristic doublet peaks of Sn $3d_{5/2}$ and Sn $3d_{3/2}$ are clearly seen around ~486 eV and ~494-495 eV, respectively, in the Sn 3d spectrum[22]. The peak positions are related to $Sn^{4+}$ chemical states, and it has been confirmed that the incorporation of Sn into the Heusler lattice has been successful. There are no peaks in the spectrum that could be attributed to the presence of any other impurities, which is another sign of the chemical purity and homogeneity of the sample. The Al 2p spectrum shows a wide peak at the equivalent of a core level at ~74 eV. The observed binding energy is typical to the $Al^{3+}$ state and a surface oxidation of Al[23]. The increase in broadening of the peak could also be attributed to a change in the local electronic environment caused by the substitution of Al at the Sn position in the Heusler structure.

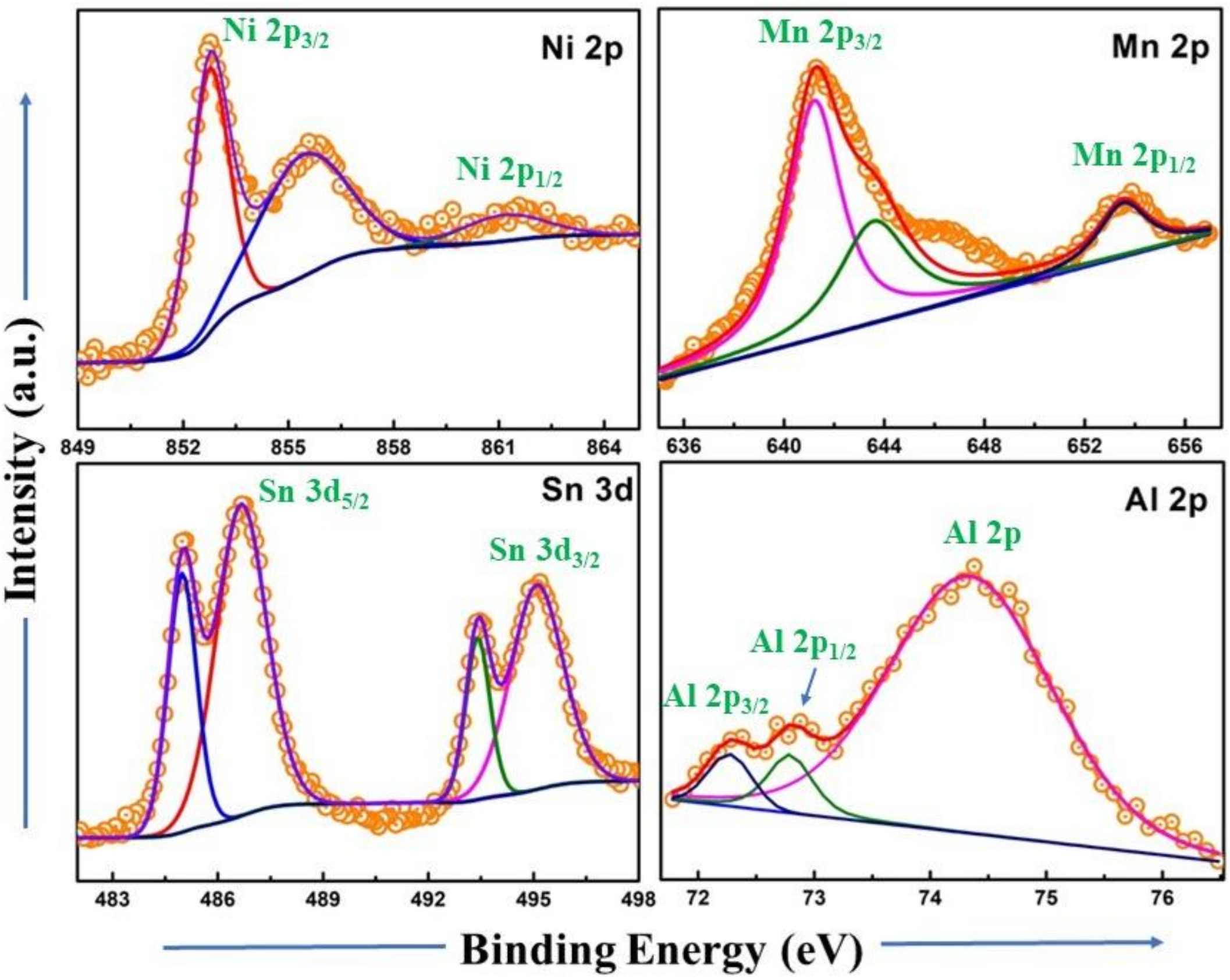

**Figure 3 :** XPS spectra of $Ni_2MnSn_{0.75}Al_{0.25}$Heusler alloy

**D.C. Magnetization Study:**

Temperature-dependent DC magnetization curves were taken using zero-field-cooled (ZFC) and field-cooled warming (FCW) at applied fields of 200 and 1000 Oe. As shown in fig. 4(a) Bifurcation is present between ZFC and FCC, suggesting that a glassy state might be present in the system.[24] In case of an applied magnetic field of 0.1 T, the ZFC magnetization curve shows a sudden increase with decreasing temperature, marked by an abrupt jump at T=734 K, corresponding to the paramagnetic-to-weak-ferromagnetic transition as shown in Fig. 4(b). This transition corresponds to the paramagnetic-to-weak-ferromagnetic transition. The value of this sharp transition at temperature $T_C$ =734K was confirmed from the first order derivative of the ZFC curve versus temperature, as shown in inset of Figure 4(b).In Ni–Mn–X based Heusler alloys a diffusionless martensitic transformation from the cubic austenite phase is common.[25-26] Upon further reduction in temperature, magnetization remains nearly constant until 263K ($T_M$); subsequent decreases in temperature result in an increase in magnetization from 3.7 emu/gm to 14.8 emu/gm.The increase in magnetization is most likely related to the martensitic phase transformation, which modifies the magnetic exchange interactions within the alloy.This structural transformation modifies the crystal lattice, thereby altering the Mn-Mn interatomic distances and magnetic exchange interactions.[27]. Due to this structural transition, the crystal lattice changes, which in turn alters the interatomic distances, especially between Mn atoms, and so do the magnetic exchange couplings. In cubic austenite, some Mn atoms may be close enough to have antiferromagnetic exchange (especially Mn atoms on the Z site in Ni-Mn-Z alloys). The Mn(Y)-Mn(Z) (B2 type disorder) distance in the cubic phase can be short enough for direct or superexchange antiferromagnetic (AFM) coupling to occur [28]. During the martensitic transition, the structure distorts, and certain Mn-Mn distances increase, weakening AFM coupling and allowing ferromagnetic (FM) exchange to dominate. This trend is also reflected in band-structure calculations of Heusler alloys [29]. This weakening facilitates the dominance of ferromagnetic exchange interactions mediated by Ni atoms. This can increase the net magnetic moment, even in a less symmetric lattice.

Furthermore, the Curie-Weiss temperature ($\theta_{\mathrm{CW}}$) and the effective magnetic moment ($\mu_{\mathrm{eff}}$) were determined by fitting the high-temperature inverse susceptibility ($\chi^{-1}$)data using the conventional Curie-Weiss relation Eq. (1):

$$\chi^{-1} = \frac{\mathrm{H}}{\mathrm{M}} = \frac{3\mathrm{k_B}}{\mu_{\mathrm{eff}}^2}(\mathrm{T} - \theta_{\mathrm{cw}}) \qquad (1)$$

The fitting is performed over the elevated-temperature region$T > 757$K, as shown in Fig.3(c). The fitting yielded the following values: Curie-Weiss temperature ($\theta_{\mathrm{CW}}$) of 746.4 K and an effective magnetic moment ($\mu_{\text{eff}}$) of 6.82 $\mu_B$. The positive value of $\theta_{\text{CW}}$indicates the predominance of ferromagnetic ordering in the alloy.

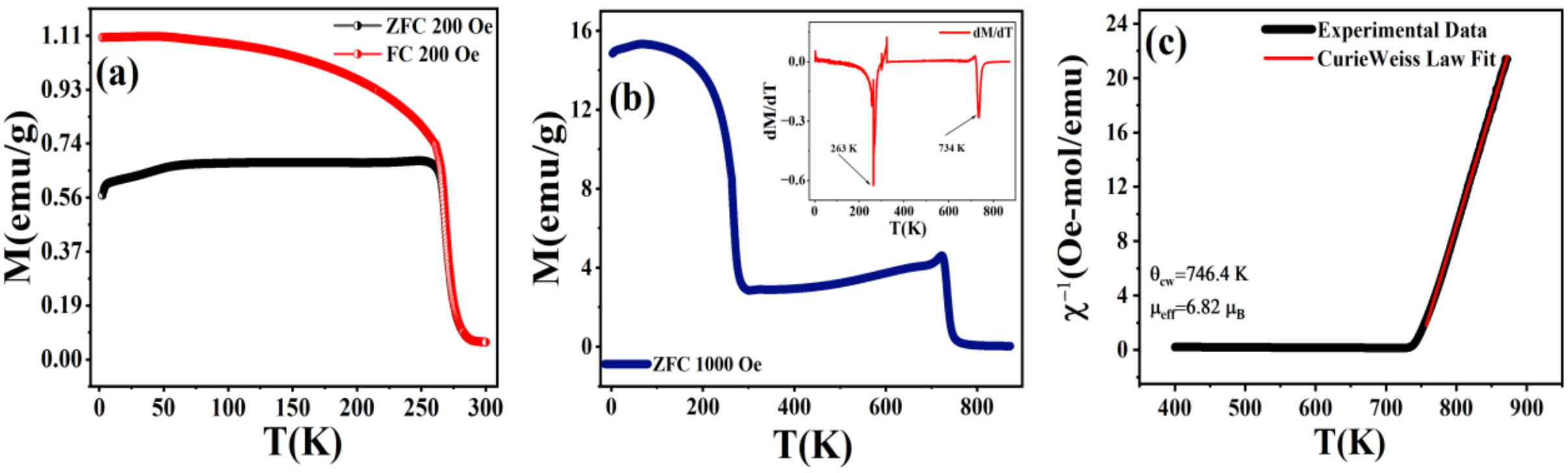

Figure 4:(a) ZFC and FC Temperature-dependent magnetization curve at 200 Oe. (b)Temperature-dependent magnetization curve at 1000 Oe in the temperature range 2K - 872K, with an inset illustrating the derivative of the magnetization with respect to temperature (c). Inverse DC susceptibility versus temperature at 200 Oe.

For better understanding of the magnetic interaction, we measured the isothermal magnetization M($\mu_0$H) of the system at various temperature values, as illustrated in Figures 5(a) and 5(b). The M($\mu_0$H) curves recorded at various temperatures (2 K, 25 K, 100 K, and 300 K) show a rapid increase in moment at low field regions; however, even at high magnetic fields up to 4 T, the magnetization does not attain complete saturation, indicating antiferromagnetic ordering Figure 5(a). Also, the loops in these M($\mu_0$H) curves exhibit clear hysteresis with a finite coercivity, indicating the presence ofFM ordering. This behaviour is typically associated with competing magnetic interactions, such as coexisting FM and AFM exchange among Mn atoms. At elevated temperatures shown in Figure 5(b), a well-defined hysteresis loop is still observed at 673 K, indicating the FM ordering persists in this regime; however, at 773 K, no hysteresis is observed, suggesting the system is in the paramagnetic region. All the M-$\mu_0$H curves indicate the presence of mixed interactions in the system. The anisotropy in the system is due to B2-type disorder, as observed in the XRD data. In the case of the regular $L_{21}$ structure, the Mn atom shares the FM interaction between them, while due to the partial occupancy of Mn at the Sn site as in $B_2$ type disorder, it offers an AFM interaction between the Mn atom and this is clear from the magnetic data.[7]The Mn-Mn spacing in the B2 type disorder is smaller than in the $L_{21}$ phase. In the $L_{21}$ phase of NMSA, the shortest Mn-Mn distance is approximately 0.411 nm. However, in the $B_2$ phase, where Mn and Al atoms are randomly distributed, the shortest Mn-Mn separation can be reduced significantly to around 0.291 nm. Sometimes, systems that exhibit anisotropy are potential candidates for applications such as the exchange bias effect and spin-valve-like effects.

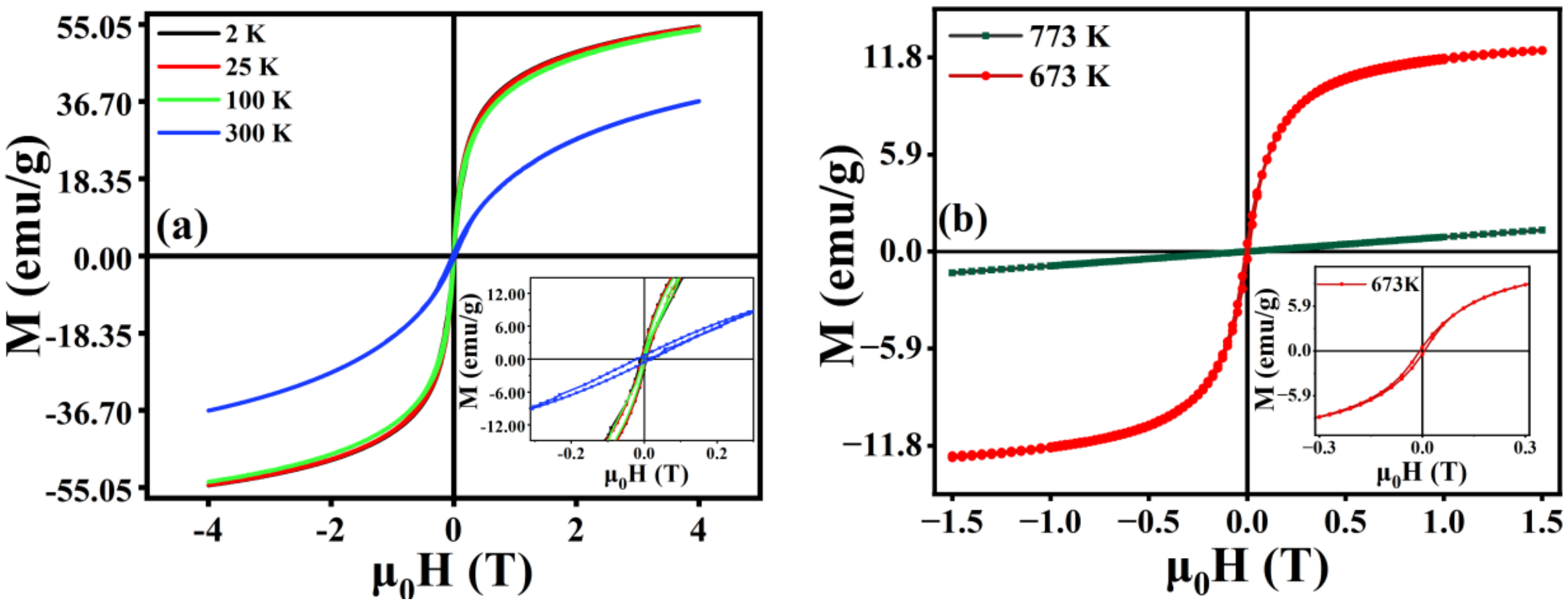

Figure 5: M-H curves at different temperatures: (a) 2–300 K and (b) 673–773 K, with insets showing low-field hysteresis.

## Differential Scanning Calorimetry:

To further corroborate the structural and magnetic transitions identified from dc magnetization measurements. The differential scanning calorimetry (DSC) measurements were carried out over the 200-900 K temperature interval. The DSC curve of NMSA is shown in Figure 6. Two distinct thermal anomalies are evident in the heat-flow curve. The first anomaly appears as an endothermic peak located at approximately 272 K, which closely corresponds to the sharp increase in magnetization occurring near $T_{\mathrm{M}} \approx 263\mathrm{K}$ in the temperature-dependent magnetization $M(T)$ data. The evolution of latent heat at this temperature unambiguously verifies that the transition is of first-order nature, a defining feature of the martensitic transformation, involving a diffusionless and displacive structural change from the high-symmetry cubic austenite phase into the low-symmetry martensitic phase [30]. The small difference of ~9 K between the DSC (272 K) and M(T) (263 K) results is attributed to the fact that the transition temperature measured on heating is typically higher than that on cooling.

The second anomaly in the DSC curve is observed near T≈733 K, which corresponds well with the Curie temperature $T_c \approx 734$ K independently identified from the minimum of dM/dT in the M(T) data measurements.[10]

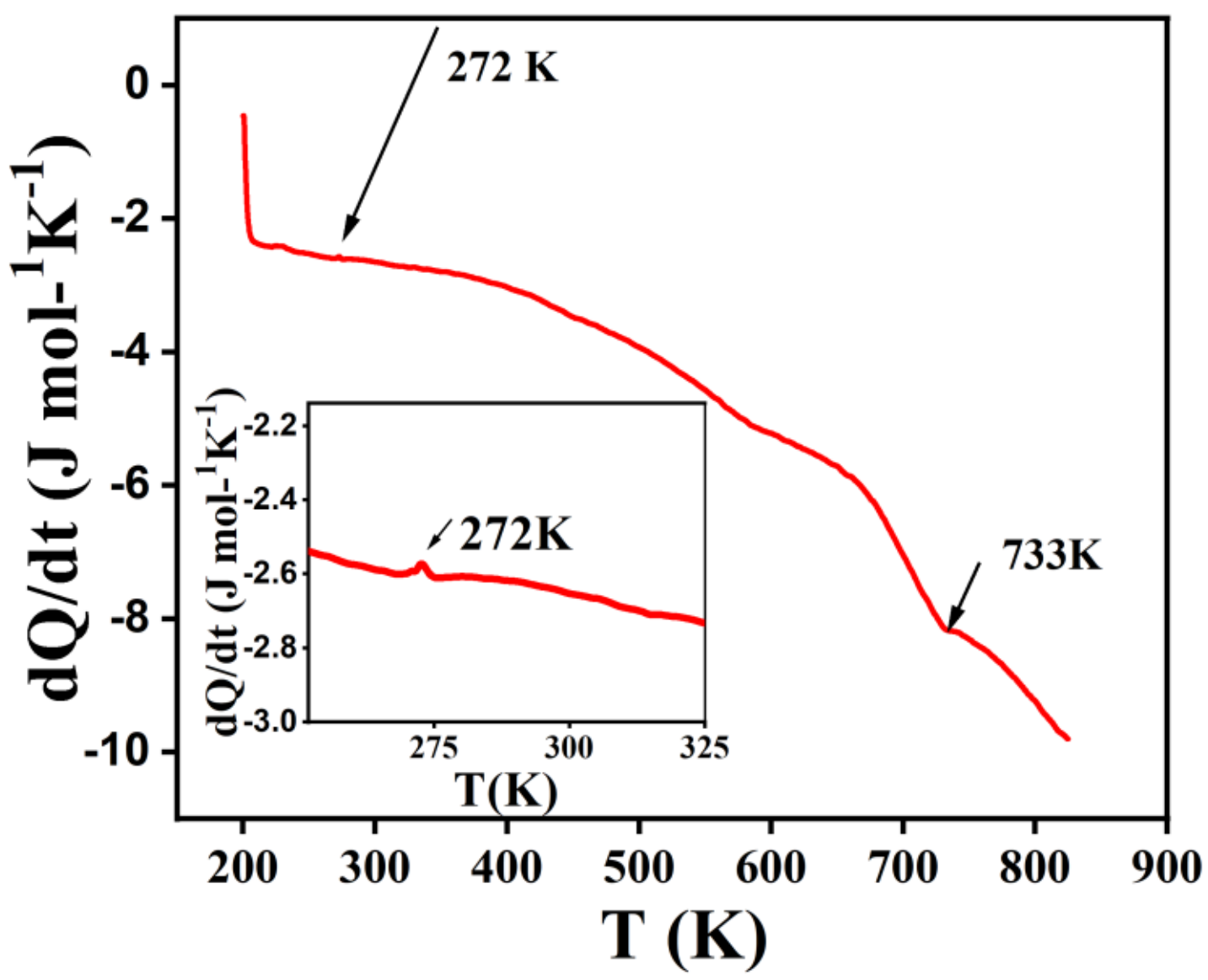

Figure 6.dQ/dT vs temperature for the alloy, showing transitions at ~272 K and ~733 K (inset: zoom near 263 K).

## AC Susceptibility Measurement:

AC susceptibility measurements serve as a remarkably sensitive tool for probing spin dynamics, thereby proving an effective method for investigating glassy magnetic behavior. To investigate the transition characteristics and scrutinize spin-glass dynamics, the sample was cooled in a zero field. Subsequently, AC susceptibility was measured at various frequencies under a constant magnetic field of $H_{ac} = 3$ Oe.To verify this glassybehaviorobserved in M(T) curves, the temperature dependence of the real and imaginary components of the AC susceptibility was measured from 6 K to 300 K across a discrete frequency range of 97 Hz to 748 Hz, as illustrated in Figures 7(a) and 7(b). The clear peaks in the imaginary part of the ac susceptibility ($\chi''_{ac}$) are noted near 39 K. The observed peaks are broad and show a clear frequency dependence, suggesting magnetic frustration and slow spin-relaxation processes which are characteristic of the glassy state.[31] These peaks also reflect the presence of underlying slow-spin relaxation processes characteristic of the glassy state.[32] However, the low-temperature spin glass phase deviates from its conventional behavior and is better described as a reentrant spin glass (RSG) state.[33]-37] In RSG systems, long-range FM, AFM, or a combination of both magnetic orderings generally emerges above the spin-glass transition. RSG behavior has been reported in a wide variety of materials, including oxides, heuslers, rare-earth intermetallic compounds, and shape-memory alloys. In the majority of reported RSG systems, the high-temperature phase exhibits ferromagnetic ordering. We have conducted a thorough analysis of the data employing multiple models to achieve a deeper understanding of the RSG state. First, the Mydosh parameter ($p$) is evaluated. It is defined as the relative variation in the freezing temperature ($T_f$) corresponding to a one-decade increase in the measurement frequency, which serves as an important criterion for distinguishing between SG and superparamagnetic (SPM) systems.[38]

$$p = \frac{\Delta T_f}{T_f \, \Delta(log_{10} f)} \tag{2}$$

where $T_f$ is freezing temperature at frequency f ;$\Delta T_f = T_{f1} - T_{f2}$ and $\Delta log_{10}(f) = log_{10}(f_1) - log_{10}(f_2)$. The parameter $p$ typically falls within the range of 0.005-0.08 for SG or cluster-glass (CG) systems, whereas for superparamagnetic systems, its value is generally greater than 0.2. To determine the Mydosh parameter, two outermost frequencies, $f_1$= 97 Hz and $f_2$ = 748 Hz, were considered. For the present system, this parameter is found to be $p = 0.0414$. It is approximately an order of magnitude higher than the reported values for SG systems, including CuMn ($p = 0.005$) and AuMn ($p = 0.0045$) [32,33], while being about an order of magnitude lower than those typically associated with superparamagnetic systems (for instance, the ideal noninteracting system α-[$Ho_2O_3(B_2O_3)$]with $p \approx 0.28$) [38].Notably, this value falls within the range commonly attributed to CG systems, thereby classifying the present system as a CG [35-37]. The spin-glass state observed in the present alloy most likely arises from the simultaneous presence of ferromagnetic (FM) and antiferromagnetic (AFM) exchange interactions. Their mutual competition creates a frustrated magnetic landscape, suppressing the development of coherent long-range magnetic order.

Moreover, below the freezing temperature $T_f$, the spin dynamics are significantly slower in SG or CG systems. This critical slowing down near $T_f$ can be analysed using the dynamic scaling law.[32], [43]

$$\tau = \tau_0 \left( \frac{T_f}{T_g} - 1 \right)^{-z\nu} \quad (3)$$

Where the relaxation time ($\tau$) represents the characteristic timescale of the dynamic fluctuations and is related to the experimental observation time through $t_{\mathrm{obs}} = 1/(2\pi\nu)$. The parameter $\tau_0$ corresponds to the microscopic relaxation time of the fluctuating entities associated with a single spin-flip process, whereas $T_g$ represents the equilibrium spin-freezing temperature in the zero-frequency ($\nu \rightarrow 0$) limit. The parameter $z\nu$ denotes the dynamic critical exponent.

To analyse the data more carefully, the power law presented in Eq. (3) may be further reformulated as follows:

$$ln(\tau) = ln(\tau_0) - z\nu \; ln\left( \frac{T_f}{T_g} - 1 \right) \quad (4)$$

The variation of $\ln(\tau)$ as a function of $ln\left( \frac{T_f - T_g}{T_g} \right)$ was fitted using the critical scaling law, as shown in Figure 8(a). The spin-freezing temperature, $T_g$, was obtained by extrapolating the $\tau$–$T_f$ relation to $f = 0$, yielding $T_g = 37.6$K. The fitting yielded $\tau_0 = 1.678 \times 10^{-8}$s, $T_g = 37.6$K, and $z\nu = 4.01$. The obtained $z\nu$ value falls within the typical range reported for spin-glass systems (4-12), whereas the comparatively large $\tau_0$ value, relative to conventional spin glasses ($\sim 10^{-12}$–$10^{-13}$ s), suggests that the magnetic freezing originates from finite-sized spin clusters having a slower relaxation, in comparison to individual spins [53-54].

The existence of interacting clusters is further supported by the breakdown of the Arrhenius law, and this law can be written as:

$$l\,n\,\tau = l\,n\,\tau^* + \frac{E_a}{k_B\,T} \quad (5)$$

This model does not agree well with spin dynamics, as illustrated in Figure 8(b). The parameters extracted from the fitting are physically unreasonable. This model is valid only for non-interacting or relatively weakly interacting clusters; therefore, its inadequacy here indicates the presence of interacting clusters inside the spin glass regime.

Here, $\tau^*$has asimilar physical significance to that of $\tau_0$, whereas $E_a/k_B$corresponds to the average activation energy barrier governing the relaxation process. The values of $\tau^*$and $E_a/k_B$were determined by fitting the $\ln(\tau)$versus $\frac{1}{T_f}$data measured in the absence of an applied magnetic field. The fitting yielded $\tau^* \approx 1.455 \times 10^{-27}$s and $E_a/k_B \approx 2197.37$K, both of which are physically unrealistic. The physically unrealistic fitting suggest that the observed magnetic dynamics do not originate from independent single-spin flips but rather from the cooperative behavior of spin clusters[41-43].

Furthermore, the presence of RSG at low temperature is also investigated by another phenomenological law, i.e., Vogel-Fulcher law, which considers the interaction among spins.[48]

$$\tau = \tau_0 \, exp\left(-\frac{E_A}{K_B(T_f - T_0)}\right) \tag{6}$$

Here, $T_0$denotes the empirical Vogel–Fulcher temperature, which provides an estimate of the interaction strength between the dynamic magnetic entities. For the fitting analysis, the equation is rewritten Eq.(6) as follows.

$$ln\,\tau = ln\,\tau_0 + \frac{E_a}{k_B(T_f - T_0)} \tag{7}$$

We plotted $\ln(\tau)$versus $\frac{1}{T_f - T_0}$,which was well fitted using Eq. (7). The corresponding fitting curve is presented in Figure 8(c). The fitting yielded the parameters $\tau_0 \approx 10^{-8}$s, $T_0 = 34.5$K, and $C = 51.12$ K. The relatively large value of $\tau_0$is inconsistent with a conventional spin-glass (SG) state and instead supports the presence of a cluster-glass (CG) state at low temperatures.

Also, a nonzero value of $T_0$ suggests the presence of finite spin–spin interactions and, consequently, supports the formation of clusters. Furthermore, within the framework of the Vogel-Fulcher model, the ratio of $T_0$ $and$ $E_a/k_B$provides insight into the coupling strength, smaller values correspond to weak coupling, while larger values indicate strong coupling. In the present case, $T_0 \approx 0.674 \left(\frac{E_a}{k_B}\right)$ which places the system in an intermediate regime and implies a finite level of interaction among the magnetic entities.

Moreover, the Tholence parameter $\delta T_{\mathrm{Th}}$, derived within the framework of the Vogel-Fulcher analysis, is commonly employed to distinguish the nature of spin interactions. It is here characterized as the relative shift between the freezing temperature $T_f$ and the Vogel-Fulcher temperature $T_0$, expressed as:

$$\delta T_{\mathrm{Th}} = \frac{T_f - T_0}{T_f} \quad (8)$$

In general, the value of $\delta T_{\mathrm{Th}}$ approaches 1 for systems of non-interacting particles, as observed in superparamagnetic (SPM) systems. In the present case, the obtained Tholence parameter $\delta T_{\mathrm{Th}} = 0.13$ lies within the range typically associated with interacting spin systems, thereby confirming the existence of a glassy magnetic state.

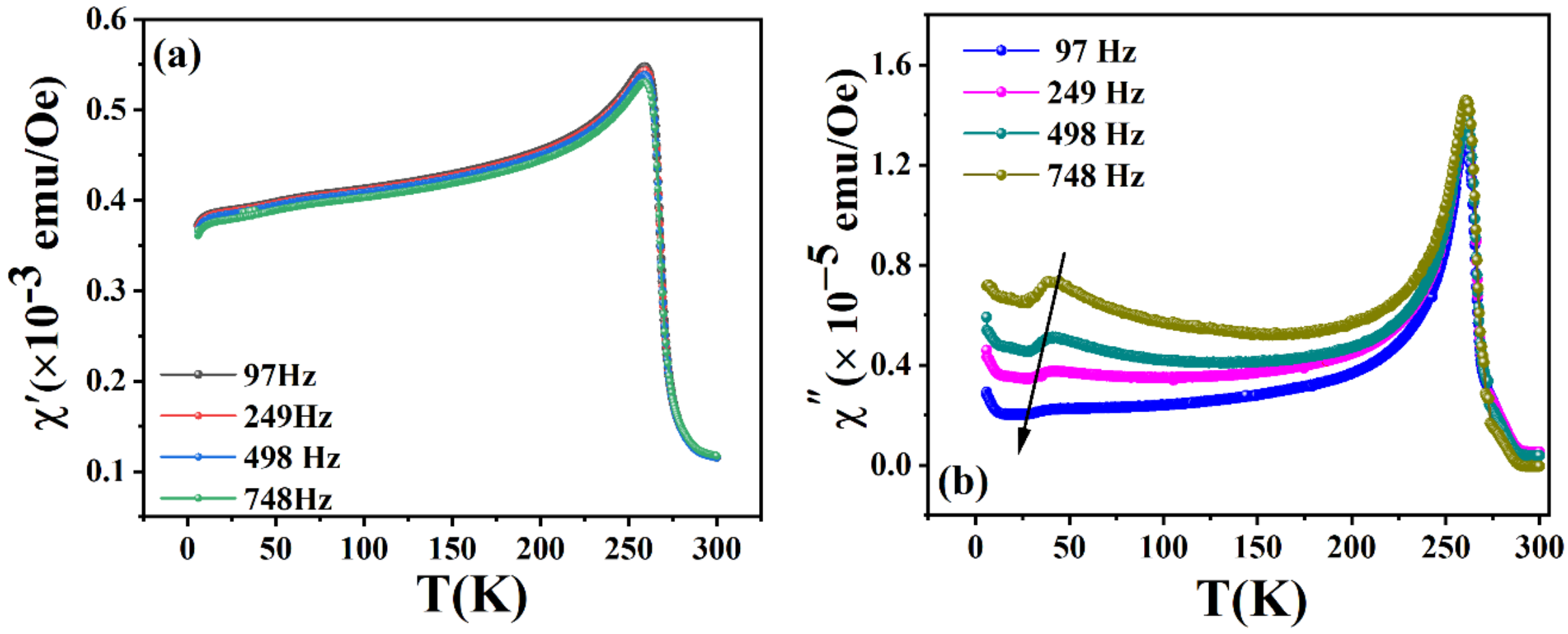

**Figure 7.** AC magnetic susceptibility of NMSA as a function of temperature measured at excitation frequencies between 97 and 748 Hz: (a) real ($\chi'$) and (b) imaginary ($\chi''$) susceptibility components.

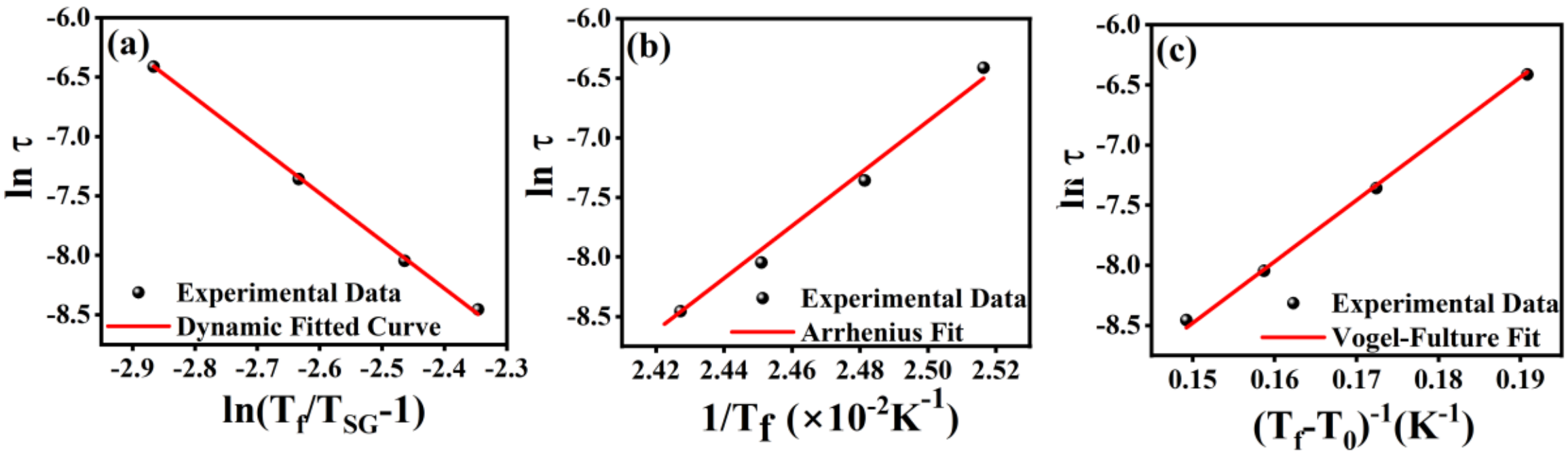

**Figure 8.** Frequency-dependent spin dynamics analysis of NMSA: (a) dynamic scaling fit, (b) Arrhenius fit, and (c) Vogel-Fulcher fit of the AC susceptibility data.

## Relaxation Effect:

The slow spin relaxation behavior in NMSA was further investigated by monitoring the time evolution of the remanent magnetization, $m(t)$, measured at temperatures below the CG freezing temperature. For this measurement, the NMSA system was initially cooled from 300 K to 25 K (below $T_f$) under an applied magnetic field of 100 Oe. After switching off the applied field, the system was left undisturbed for 100 s. The magnetization was then measured for 6000 s to observe its relaxation. Fig. 9 presents the variation of the normalized remanent magnetization with time, recorded at 25 K in the CG regime.It clearly suggests that even after a long duration, the residual magnetization persists to decay, which is a characteristic feature of disordered glassy systems arising from the presence of hierarchically arranged metastable states.[39,45]. In disordered systems, this kind of isothermal relaxation behavior is typically analyzed using the phenomenological Kohlrausch-Williams-Watts (KWW) stretched - exponential model, expressed as follows [49]

$$m(t) = m_0 - m_g \; exp\left\{-\left(\frac{t}{\tau}\right)^{\beta}\right\} \tag{9}$$

In this context, $m_g$corresponds to the magnetization associated with the glassy component, where $m_0$denotes the initial remanent magnetization, $\tau$ represents the characteristic relaxation time, and $\beta$ is the stretching exponent (shape parameter) which typically takes values between 0 and 1 in SG systems.

However, the experimental data could not be adequately described using the KWW model.As shown in the inset of Figure 9, the data deviate clearly from the KWW model in the rapidly decaying region ($t < 240$ s), whereas a reasonably good fit is obtained at longer times ($t > 240$s).

For systems consisting of interacting magnetic particles, the time evolution of spin relaxation can be analyzed through the relaxation rate, W(t), defined as $W(t) = \frac{d[\ln m(t)]}{dt}$. According to the model proposed by Ulrich and co-workers, this relaxation rate exhibits a power-law dependence on time and can be expressed as $W(t) = At^n$. Here, $A$ is a temperature-dependent prefactor, while $n$ is a critical exponent associated with the spin density and the magnetic interaction strength[50].In this model**,** the residual magnetization can be described by the following expression, as proposed by Ulrich *et al. al* [47,48]

$$m(t) = \frac{M(t)}{M(t_0)} = e^{-\alpha_n}\left[1 + \alpha_n\left(\frac{t}{t_0}\right)^{1-n}\right] \quad (10)$$

Here,$\alpha_n = \frac{A\, t_0^{1-n}}{|1-n|}$ and $M(t_0)$denotes the initial magnetization at time $t_0$.

A satisfactory fit was obtained when the above equation was applied to the experimental results, confirming the model's validity (Figure 9). Moreover, the obtained exponent value of (n = 1.03) is characteristic of a cluster glass state, indicating the presence of glassy magnetic dynamics in the system.[53]

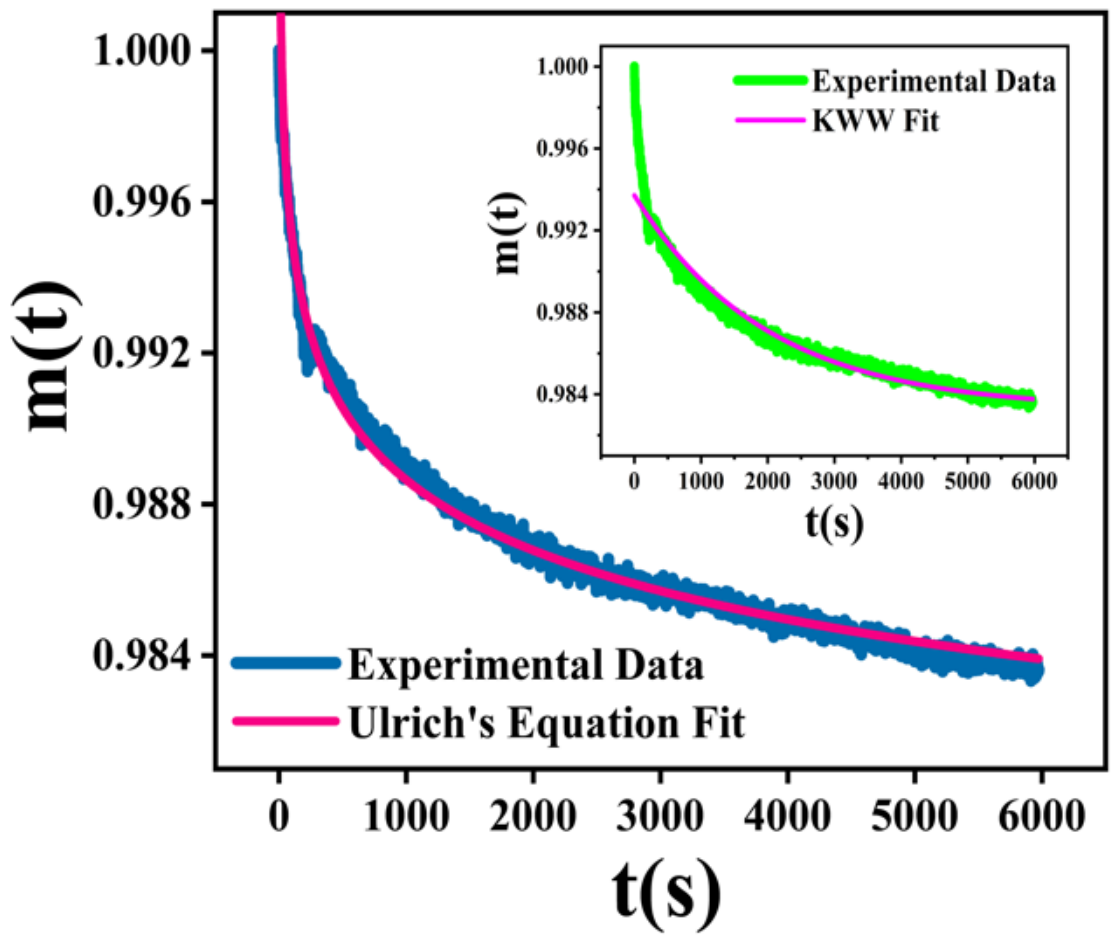

Figure 9: Time-dependent thermoremanent magnetization of NMSA fitted using Ulrich's equation. The inset shows the stretched exponential (KWW) fitting of the experimental data.

## AC magnetic susceptibility (H):

Figure 10(a) shows the temperature dependence of χ″(T) for the NMSA system measured under various applied dc magnetic fields. It is clear that the χ″ peaks shift to lower temperatures with increasing dc magnetic field, particularly around 39.4 K ($T_f$), consistent with the characteristic behavior of a glassy magnetic state.To investigate the effect of an external magnetic field on the freezing behavior, the low-field $T_f(H)$data were fitted using the de Almeida-Thouless (AT) model. This relation is commonly applied to describe the field dependence of the freezing temperature in SG and CG systems and is given by Eq. (11).

$$H_{dc}^{2/3} = \Delta J^{2/3}\left(1 - \frac{T_f(H)}{T_f(0)}\right) \quad (11)$$

The parameter $T_f(0)$denotes the freezing temperature measured in the absence of an external dc magnetic field, while ΔJ reflects the magnitude of the magnetic interactions within the system.As seen in Fig.10 (b), the experimental $T_f(H)$data are well represented by the de Almeida-Thouless model. Extrapolation of the fitted curve to zero magnetic field gives $T_f(0)$ to be ~42.79 K, which is nearly identical to the freezing temperature obtained from the temperature-dependent AC susceptibility measurements at different frequencies.

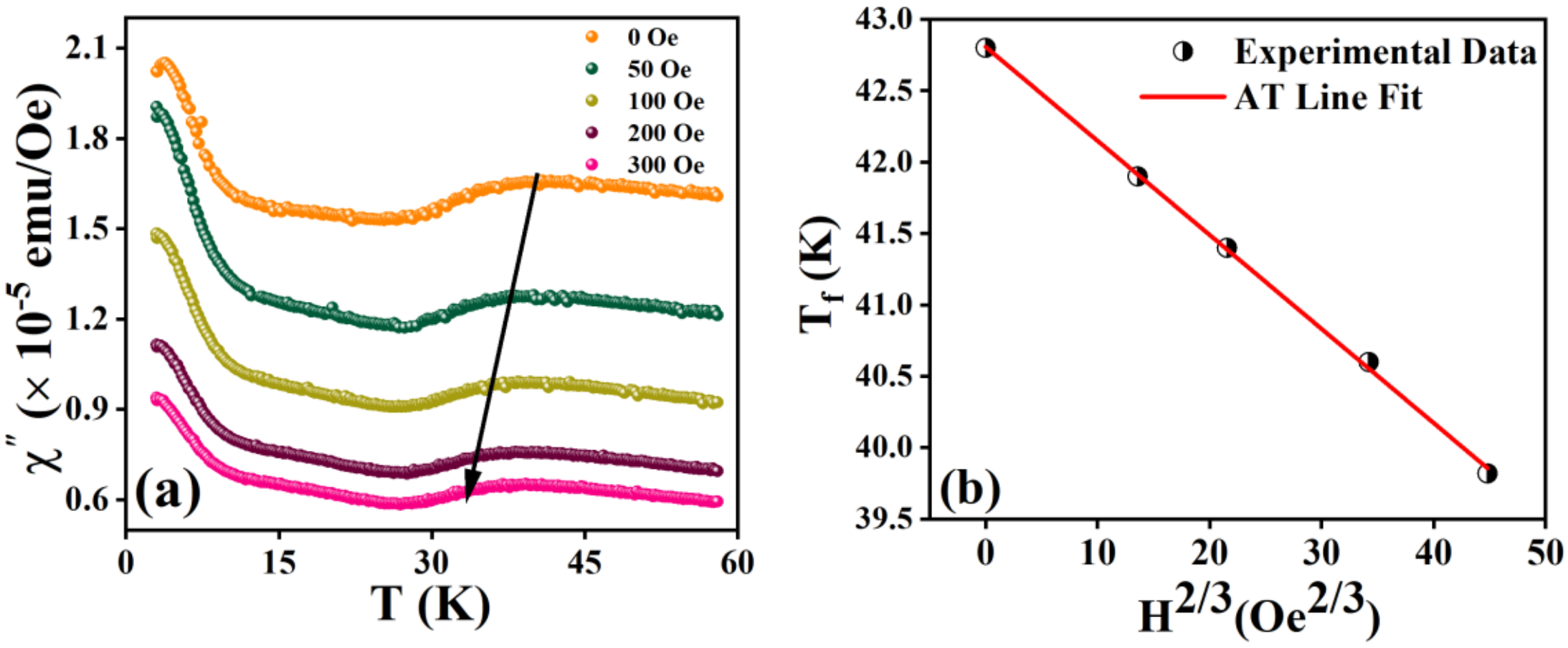

Figure 10(a): Temperature dependence of the imaginary part of AC susceptibility $(\chi'')$of NMSA under different DC magnetic fields. (b) Almeida-Thouless (AT) line fitting of freezing temperature $T_f$as a function of $H^{2/3}$

## Magnetocaloric Effect:

The magnetocaloric effect (MCE) refers to the change in magnetic entropy of a material when an external magnetic field is applied under either adiabatic or isothermal conditions.For this isothermal magnetization data were recorded up to magnetic fields of 5T across the temperature intervals associated with the observed magnetic transitions. The correspondingM-$\mu_0$Hisotherms are displayed in Figures 11(a) and 11(b). Maxwell's thermodynamic relation was used to evaluate the change in the isothermal magnetic entropy, as given by the equation below:

$$\frac{\partial S_M(T,H)}{\partial H} = \mu_0 T \frac{\partial M(T,H)}{\partial T} \qquad \textbf{(12)}$$

Using this relation, the magnetic entropy change, $\Delta S_M$ (T,H) of the material can be determined. [54]

$$\Delta S_M(T,H) = \mu_0 \int \frac{\partial M}{\partial T} dH \qquad \textbf{(13)}$$

To facilitate the calculation from experimentally acquired (M(H, T)) data, the above relation was reformulated into the following discrete form:[55]

$$\Delta S_M(T,H) = \mu_0 \sum_{n,n+1} \left[\frac{M_n(T_n,H) - M_{n+1}(T_{n+1},H)}{T_{n+1} - T_n} \times \Delta H\right] \qquad (14)$$

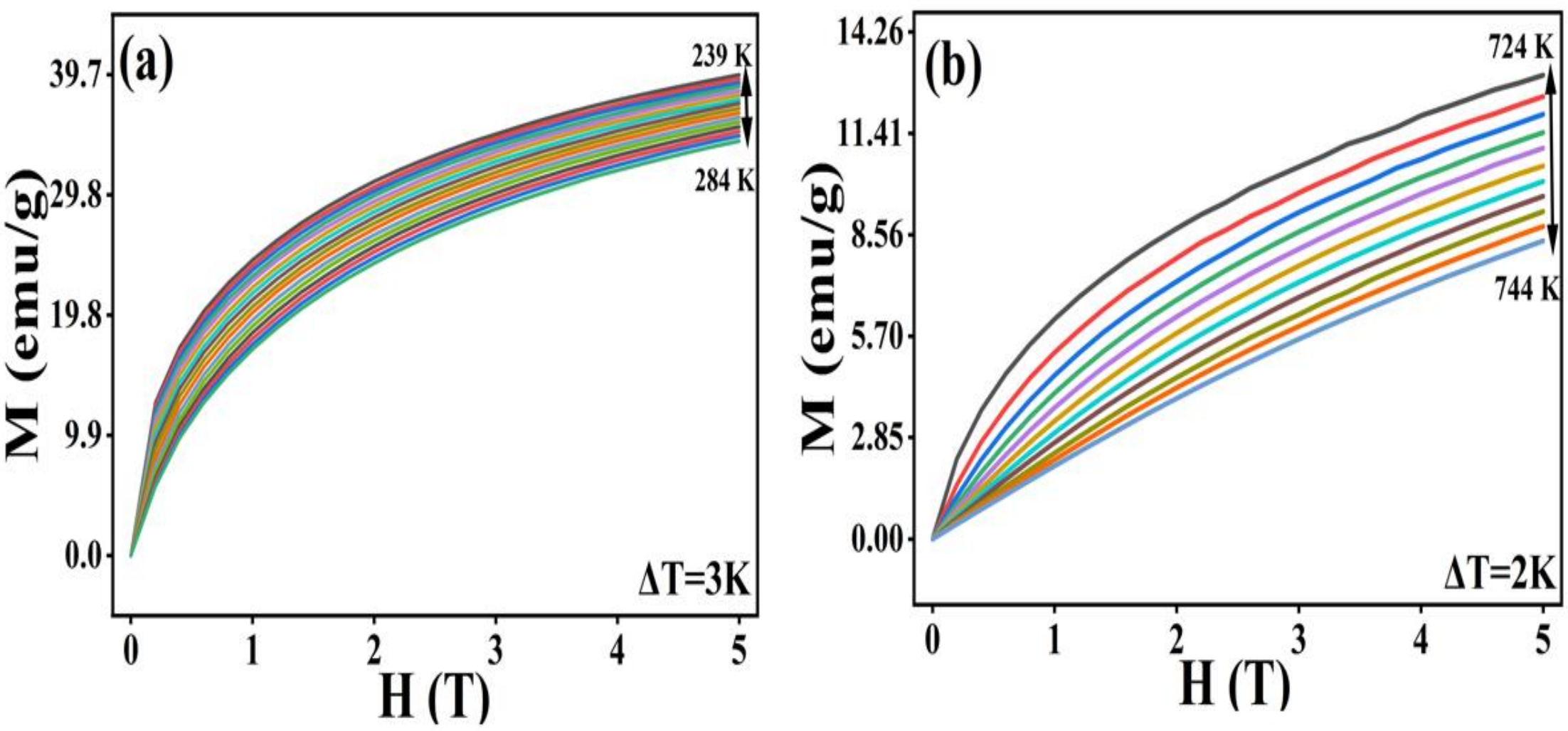

Figure11: Isothermal magnetization as a function of applied magnetic field up to 5 T in the temperature range (a)239K-284K with $\Delta$ T= 3K, (b)724K-744 K with $\Delta$ T= 2K.

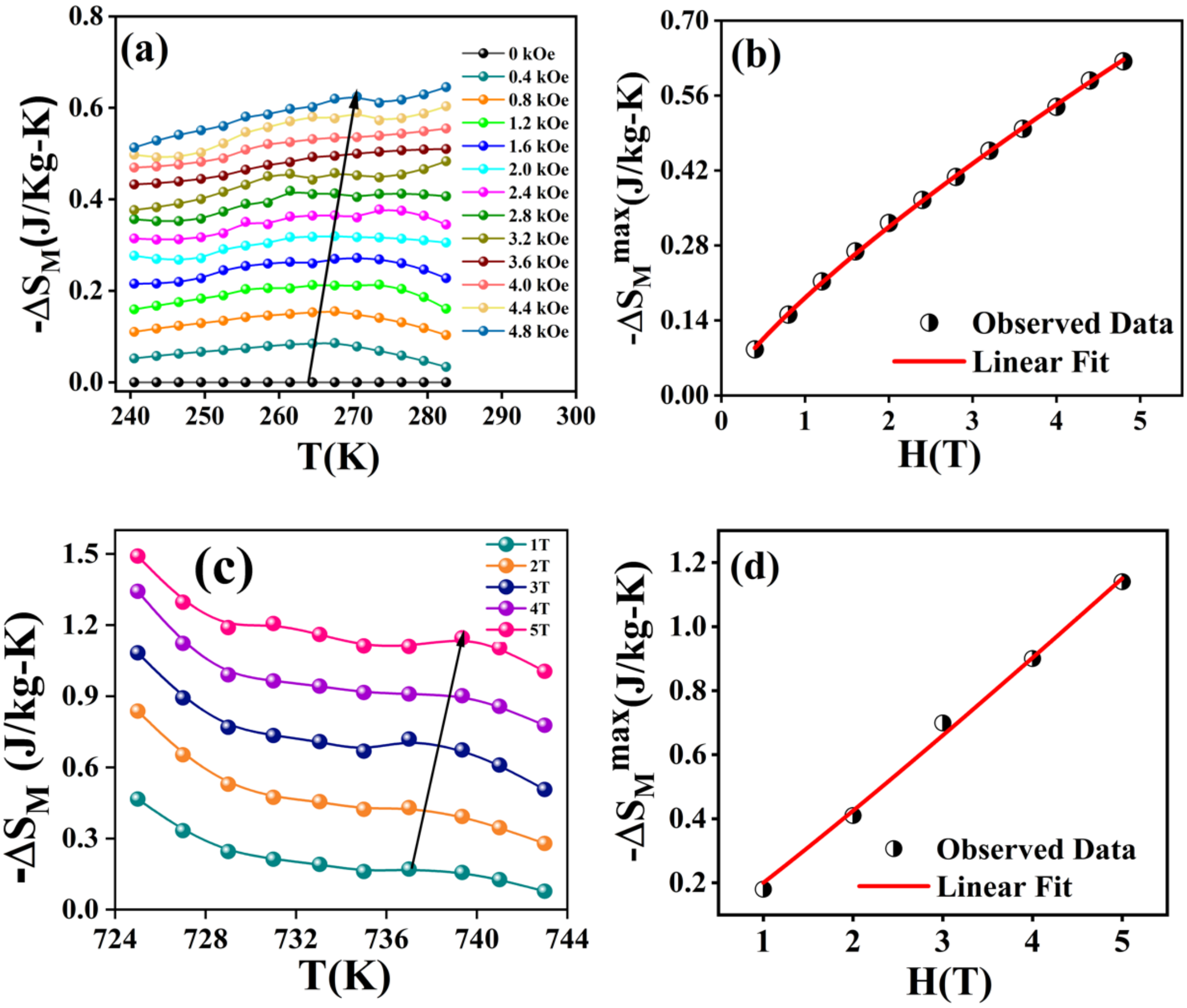

Figure 12. (a) and (c) Temperature dependence of the magnetic entropy change ($\Delta S_M$) of NMSA under magnetic field from 0T to 5T in temperature range 239K-284K and 724 K-744 K respectively. (b) and (d) Variation of maximum magnetic entropy change ( $\Delta S_M^{Max}$) applied magnetic field, showing a power law fit $\Delta S_M \sim H^n$.

The temperature dependence of $\Delta S_M^{max}$, revealing distinct peaks near the transition at 263 K, as shown in Figure 12(a). The field dependence of $\Delta S_M$ follows a power law that can be expressed by: $\Delta S_M = aH^n$**,** where a is a constant and n is the field exponent. As shown in Figure 12(b), for NMSA, the exponent n is 0.784. It is well known that n approaches 1 in ferromagnetic phases and tends toward 2 in AFM or paramagnetic phases[51,52]. Moreover, in Figure 11(b), we also show the M-H isotherm measured from 724 K to 744 K. Near the transition region, where the system goes to a completely paramagnetic region, again, the temperature dependence of $\Delta S_M$, revealing distinct peaks near the transition temperature, power law $\Delta S_M \sim H^n$ is fitted in this region. The value of n is 1.07. The value of n confirms that the system is in the ferromagnetic phase.

## Exchange bias:

Several studies have reported a unidirectional exchange bias effect in Ni-based HAs.The exchange-bias effect is primarily governed by competing interactions at the interface between two oppositely magnetically ordered lattices (FM and AFM sublattices), arising from mixed magnetic phases. Moreover, magnetic anisotropy plays a crucial role in the exchange-bias phenomenon. Unidirectional exchange bias refers to the displacement of the FC hysteresis loop in a direction opposite to the applied cooling field relative to the ZFC virgin curve.To examine this effect, the sample was cooled from 300 K to 5 K under both ZFC and FC conditions (-50 kOe), and the magnetic field-dependent isothermal magnetization was subsequently measured as shown in Figure 13. As evident from Figure 13, the FC hysteresis loop is shifted toward the positive field direction, with an exchange-bias field of approximately 110 Oe.

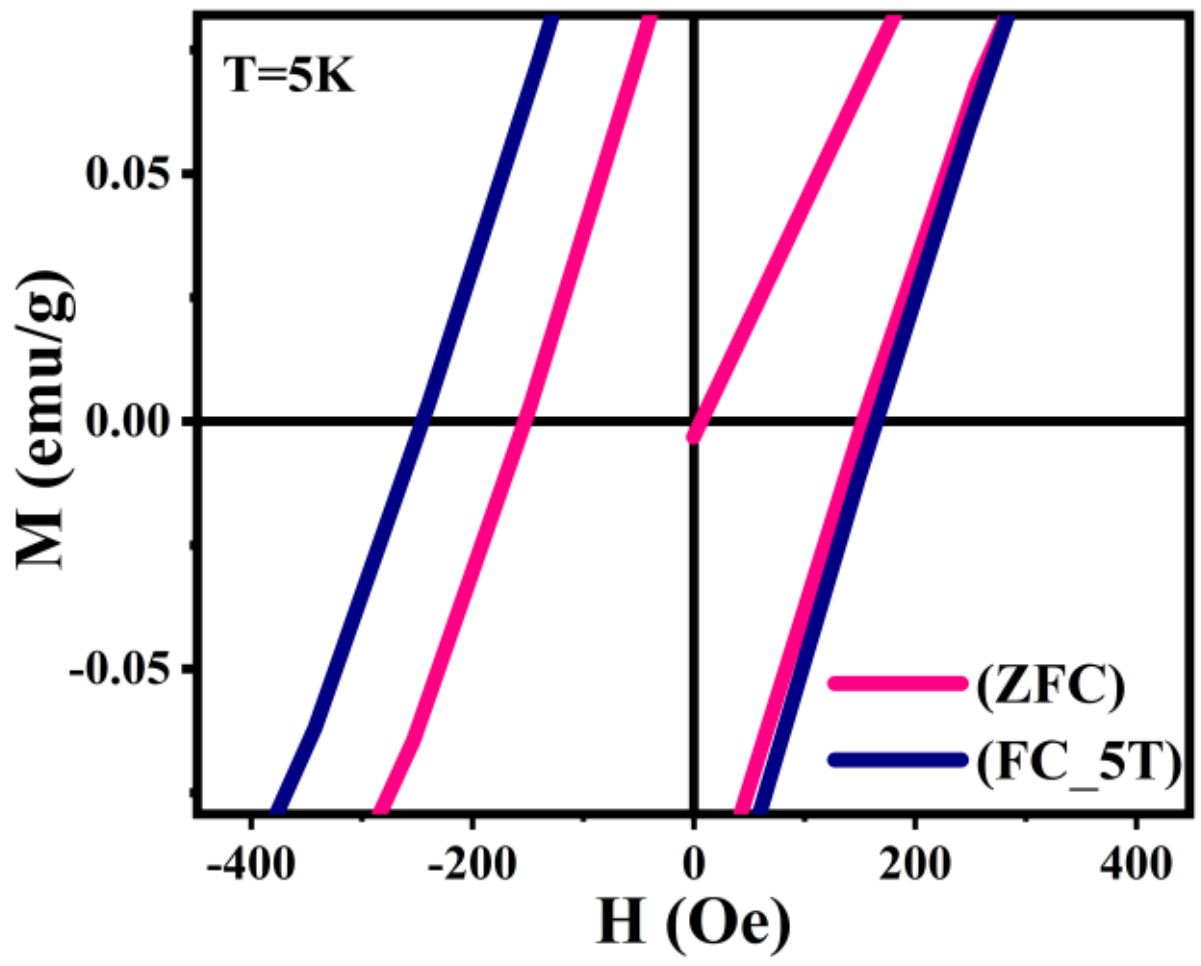

**Figure 13:** Isothermal magnetization measurements at 5 K under ZFC and FC conditions. The figure shows an enlarged view of the shift along the positive X axis upon application of a +5T field.

## Conclusion:

The present study demonstrates that Al substitution in $Ni_2MnSn$ produces a chemically homogeneous Heusler alloy with a coexisting $L2_1$–B2 crystal structure, where the structural disorder plays a crucial role in determining the magnetic ground state. The coexistence of these structural phases promotes competing ferromagnetic and antiferromagnetic exchange interactions, leading to magnetic frustration and strong magnetostructural coupling, as evidenced by the close correlation between the magnetic and martensitic transitions. The first-order nature of the martensitic transformation is further confirmed by differential scanning calorimetry, establishing a strong coupling between the structural and magnetic degrees of freedom. Temperature-dependent Raman spectroscopy further reveals pronounced phonon anomalies near ~263 K and the Curie temperature (~734 K), confirming strong spin–phonon coupling. The observed phonon renormalization demonstrates that magnetic ordering strongly influences the lattice dynamics while preserving the structural stability of the Heusler phase over the investigated temperature range.

Comprehensive magnetic investigations, including dc magnetization, frequency-dependent AC susceptibility, magnetic relaxation, and field-dependent susceptibility measurements, consistently establish the formation of an interacting reentrant cluster-glass state, rather than a conventional spin-glass or superparamagnetic state. Furthermore, the observation of an exchange-bias effect provides additional evidence for the coexistence of ferromagnetic and antiferromagnetic regions arising from the mixed magnetic interactions, highlighting the complex magnetic ground state of the alloy.

The alloy further exhibits an appreciable magnetocaloric response across the magnetic transition region, demonstrating its potential for magnetic refrigeration applications.